\documentclass[12pt,preprint]{aastex}

\newcommand{\etal  }{{et al.} }
\newcommand{\msun}{\thinspace M_\odot}

\def\lesssim{\mathrel{\hbox{\rlap{\hbox{\lower4pt\hbox{$\sim$}}}\hbox{$<$}}}}
\def\gtrsim{\mathrel{\hbox{\rlap{\hbox{\lower4pt\hbox{$\sim$}}}\hbox{$>$}}}}
\newcommand{\cm}{\,{\rm cm}^{-3} } 
\newcommand{\km}{\,{\rm km\, s}^{-1}} 
\newcommand{\nc}{n_{\rm c} } 
\newcommand{\jcm}{{\rm cm^2\,s^{-1}}}
\newcommand{\rcri}{R_{\rm c} }
\newcommand{\tc}{t_{\rm c}}

\shorttitle{Circumbinary Outflow}
\shortauthors{Machida  \etal 2009}

\begin{document}

\title{The Circumbinary Outflow: A Protostellar Outflow Driven by a Circumbinary Disk}

\author{Masahiro N. Machida\altaffilmark{1} and Shu-ichiro Inutsuka\altaffilmark{2}, and Tomoaki Matsumoto\altaffilmark{3}} 

\altaffiltext{1}{National Astronomical Observatory of Japan, Mitaka, Tokyo 181-8588, Japan; masahiro.machida@nao.ac.jp}

\altaffiltext{2}{Department of Physics, Graduate School of Science, Nagoya University, Furo-cho, Chikusa-ku, Nagoya, Aichi 464-8602, Japan; inutsuka@nagoya-u.jp}

\altaffiltext{3}{Faculty of Humanity and Environment, Hosei University, Fujimi, Chiyoda-ku, Tokyo 102-8160, Japan; matsu@i.hosei.ac.jp}

\begin{abstract}
The protostellar outflow is star's first cry at the moment of birth. 
The outflows have indispensable role in the formation of single stars, because they carry off the excess angular momentum from the centre of the shrinking gas cloud, and permits further collapse to form a star. 
On the other hand, a significant fraction of stars is supposed to be born as binaries with circumbinary disk that are frequently observed. 
Here, we investigate the evolution of a magnetized rotating cloud using three-dimensional resistive MHD nested-grid code, and show that the outflow is driven by the circumbinary disk and has an important role even in the binary formation. 
After the adiabatic core formation in the collapsing cloud core, the magnetic flux is significantly removed from the centre of the cloud by the Ohmic dissipation. 
Since this removal makes the magnetic braking ineffective, the adiabatic core continuously acquires the angular momentum to induce fragmentation and subsequent binary formation. 
The magnetic field accumulates in the circumbinary disk where the removal and accretion of magnetic field are balanced, and finally drives circumbinary outflow. 
This result explains the spectacular morphology of some specific young stellar objects such as L1551 IRS5.  
We can infer that most of the bipolar molecular outflows observed by low density tracers (i.e., CO) would correspond to circumbinary or circum-multiple outflows found in this report, since most of the young stellar objects are supposed to be binaries or multiples. 
\end{abstract}

\keywords{binaries: general---ISM: clouds---ISM: jets and outflows---ISM: magnetic fields ---MHD---stars: formation}

\section{Introduction}
Stars are born in molecular clouds which are magnetized interstellar clouds in our galaxy.
Recent observations with high-angular-resolution measurements of polarized dust emission have shown that the star formation and gas collapsing processes are certainly closely related to the magnetic field \citep{girart06,girart09}.
The existence of highly collimated outflow/jets has established an indispensable role played by the magnetic field \citep{cabrit88,cabrit07,hirano06}.
The realistic process of star formation in a magnetized molecular cloud is fully three-dimensional magneto-hydrodynamics (hereafter MHD) with self-gravity. 
Theoretical astrophysicists are now tackling this process by the method of numerical simulation thanks to the recent development of various techniques \citep{kudoh98,tomisaka02,matsu04,hennebelle08a}.  
Hydrodynamical calculations of protostellar collapse without magnetic field have shown that initially rotating molecular clouds tend to form binary or higher multiple stars \citep{goodwin07}, and the resultant binary stars are externally surrounded by circum-binary disks, which have already been observed in star-forming regions.  
On the other hand, protostellar collapse calculations with magnetic field of the observationally inferred strength show spectacular bipolar outflows \citep{tomisaka02,machida04}. 
The outflow is driven by the Lorentz force and removes the angular momentum of the driving object, which suppresses the fragmentation or the binary formation \citep{hennebelle08b,machida08b}.  
Thus, the driving of MHD outflow and the binary formation have been thought to be incompatible so far, although both the protostellar outflows and circumbinary disks have been observed numerously in star forming regions. 
Therefore, understanding of the puzzling link between them seems to require qualitative improvement in theoretical modeling. 
A clue for possible modifications are found in the evolution of magnetic field: most of the previous MHD calculations for the evolution of the collapsing cloud were done under the ideal MHD approximation that is valid only in the low-density phase of molecular gas but not in the high-density protostellar phase, which is obviously the room for the improvement for theoreticians. 
Recently we studied the star formation process in resistive MHD calculations that account for the low ionization degree in molecular clouds, and showed the magnetic field is significantly removed from the centre of the collapsing cloud and fragmentation (or binary formation) is actually possible even in a strongly magnetized cloud \citep{machida07,machida08a,machida08b}. 
In this paper, we show the dramatic relation between the binary formation and outflow in resistive MHD calculations.

\section{Model and Numerical Method}
To study the evolution of star-forming cores in a large dynamic range of spatial scale,  a three-dimensional nested grid method is used, in which the equations of resistive magnetohydrodynamics are solved (see Eq. [1]--[5] of \citealt{machida07}).
As the initial state, we take a spherical cloud with critical Bonnor--Ebert (BE) density profile, in which the uniform density is adopted outside the sphere ($r > \rcri$).
For BE density profile, we adopt the central density as $\nc =  10^{6}\cm$ and isothermal temperature as $T=10$\,K. 
For these parameters, the critical BE radius is $\rcri = 4.6\times10^3$\,AU.
To promote the contraction, we increase the density  by 1.68 times.
The mass inside $r < \rcri$ is $0.8\msun$.
The cloud rotates rigidly with $\Omega_0 = 2.7 \times 10^{-13}$\,s$^{-1}$ around the $z$-axis in the region of  $r< \rcri$, while the uniform magnetic field ($B_0 =32 $$\mu$G) parallel to the $z$-axis (or rotation axis) is adopted in the whole computational domain.
In the region of $r<\rcri$, the ratio of thermal $\alpha_0$, rotational $\beta_0$, and magnetic $\gamma_0$ to gravitational energies are $\alpha=0.5$, $\beta=0.04$, and $\gamma=0.06$, respectively.

For a realistic evolution of the magnetic field in the collapsing gas cloud, we adopted the resistivity ($\eta$) derived in \citet{machida07}.
Figure~\ref{fig:1}{\it a} shows the resistivity $\eta$ and magnetic Reynolds number $Re$ against the central number density \citep[for detailed description see \S2.2 of][]{machida07}. 
The gas temperature adopted in this study also plotted by the dotted line in this figure.

To calculate a large spatial scale, the nested grid method is adopted \citep[for details see ][]{machida05a,machida05b}. 
Each level of a rectangular grid has the same number of cells ($ 64 \times 64 \times 32 $).
The calculation is first performed with five grid levels ($l=1-5$).
The box size of the coarsest grid $l=1$ is chosen to be $2^5 \rcri$.
Thus, grid of $l=1$ has a box size of $\sim 1.5\times 10^5$\,AU.
A new finer grid is generated before the Jeans condition is violated.
The maximum level of grids is $l_{\rm max} = 13$ that has a box size of 35\,AU and the cell width of 0.54\,AU.

We calculate two models: (a) resistive and (b) ideal MHD models.
Hereafter, we call the former `resistive model', and the latter `ideal model.' 
Both models have the same initial condition shown above.
The former include resistive term in induction equation, while the latter does not.

\section{Results}
As shown in Figure~\ref{fig:1}{\it a} (the dotted line), the molecular gas obeys the isothermal equation of state with temperature of $\sim 10$ K until $n_c \simeq 10^{10}\cm$, then cloud collapses almost adiabatically ($10^{10}\cm \lesssim n_c \lesssim 10^{16}\cm$; adiabatic phase) and quasi-static core (hereafter, first core)  forms during the adiabatic phase \citep{larson69, masunaga00}.
The first core forms both in resistive and ideal MHD models when the central density reaches $n_c \simeq 2\times10^{12} \cm$, and have a radius of $\sim 10$ AU and mass of $0.017\msun$.
By coincidence, the first core formation epoch almost agrees with the epoch at which the Ohmic dissipation becomes effective (i.e., $Re<1$), as seen in Figure~\ref{fig:1}{\it a}.
Therefore, in resistive model, the magnetic field dissipates inside or around the first core with time after its formation.

The evolutions of the magnetic field $B_{z,c}$ and angular momentum $J$ after the first core formation $\tc$ both in ideal and resistive models are plotted in Figure~\ref{fig:1}{\it b}.
The figure shows that $B_{\rm z,c}$ and $J$ have the same evolution track in both models for $\tc\lesssim 100$\,yr but different tracks for $\tc\gtrsim100$\,yr.
In ideal model, the magnetic field continues to increase to reach $B_{z,c}\simeq1$\,G at $\tc \simeq 1000$\,yr, while the angular momentum decreases to have $J\simeq7\times10^{47}\,\jcm$ at the same epoch.
The decrease of the angular momentum is due to the magnetic braking which transfers the angular momentum outwardly along the magnetic field lines \citep[][]{basu94,tomisaka02}.
Since the cloud collapse slows and the rotation timescale becomes shorter than the collapsing timescale after the first core formation, magnetic field lines are strongly twisted and the angular momentum is considerably transferred owing of the amplified field.
On the other hand, in resistive model, the magnetic field begins to decrease at $\tc\simeq300$\,yr, while the angular momentum continues to increase.
Since the density of the first core exceeds $n\gtrsim10^{12}\cm$, the magnetic field is effectively removed from the first core by the Ohmic dissipation as shown in Figure~\ref{fig:1}{\it a}.
Thus, the region around and inside the first core has a weak field that little contribute to the magnetic braking, thus, the first core in resistive model has a larger angular momentum than that in ideal model.
As shown in Figure~\ref{fig:1}, the magnetic field in resistive model is about two orders of magnitude weaker than that in ideal model, while the angular momentum in resistive model is about two orders of magnitude larger that that in ideal model.

\begin{figure}
\includegraphics[width=170mm]{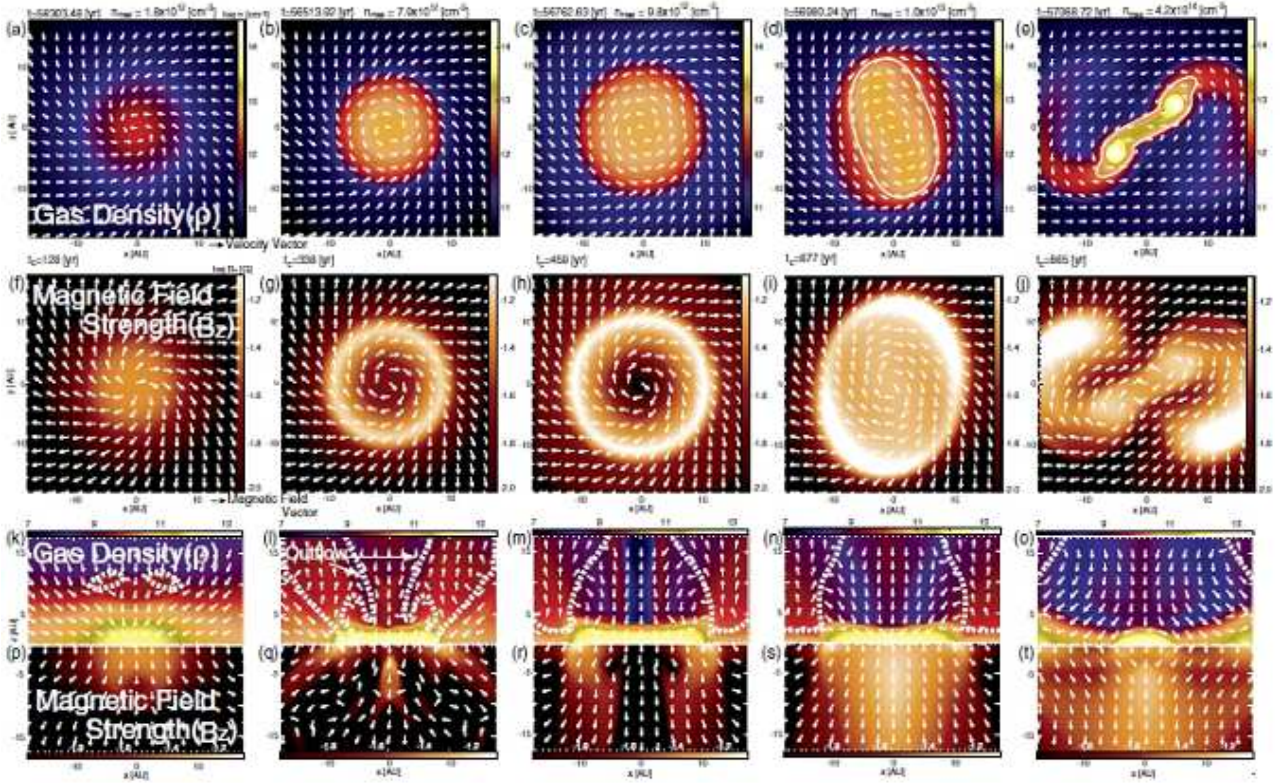}
\caption{
Distributions of density (panels {\it a}-{\it e}, and {\it k}-{\it o}) and magnetic field strength (panels {\it f}-{\it j} and {\it p}-{\it t}) around the first core (grid level $l=13$) on $z=0$ (panels {\it a}-{\it j}) and $y=0$ (panels {\it k}-{\it t}) plane.
Arrows in panels {\it a} - {\it j} mean the velocity vector, while those in panels {\it k} -{\it t} mean the magnetic field vector.
The elapsed time from the initial ($t$) and that ($\tc$) after the first core formation, and maximum number density ($n_{\rm max}$) are described in each top panel.
The dotted line in panels {\it k} -{\it o} means the border between infall and outflow inside which the gas is outflowing from the center of the cloud.
}
\label{fig:2}
\end{figure}

Figure~\ref{fig:2} shows the distribution of density and magnetic field on $z=0$  and $y=0$ planes after the first core formation.
The first core has a disk-like structure (Figs.~\ref{fig:2}{\it k}-{\it n}) with a centrally peak density profile (Figs.~\ref{fig:2}{\it a}-{\it d}).
On the other hand, the magnetic field has a peak at edge or outside of the first core: the peak of the magnetic field has a ring-like structure as shown in Figures~\ref{fig:2}{\it g}-{\it i}.
This is because the magnetic flux is removed from the region inside the first core by the Ohmic dissipation.
Figures~\ref{fig:2}{\it b}-{\it d} indicate that the first core has a density of $n=10^{12}-10^{15}\cm$ where the magnetic Reynolds number is below unity $Re<1$ (see, Fig.~\ref{fig:1}{\it a}), thus the magnetic field dissipates effectively.
Since the first core increases in size with times, the decoupled region (i.e., the region with $Re<1$) also expands.
Therefore, the peak of the magnetic field strength moves outwardly as shown in Figures~\ref{fig:2}{\it g}-{\it i}.
In resistive model, since the magnetic braking is not so effective owing to the weak field, the first core continues to spin up (see, Fig.~\ref{fig:1}{\it b}).
In general, when the first core has a large angular momentum, fragmentation occurs \citep[e.g.,][]{goodwin07}.
The fragmentation occurs $\sim800$\,yr after the first core formation (Fig.~\ref{fig:2}{\it e}) only in resistive model.

\begin{figure}
\includegraphics[width=170mm]{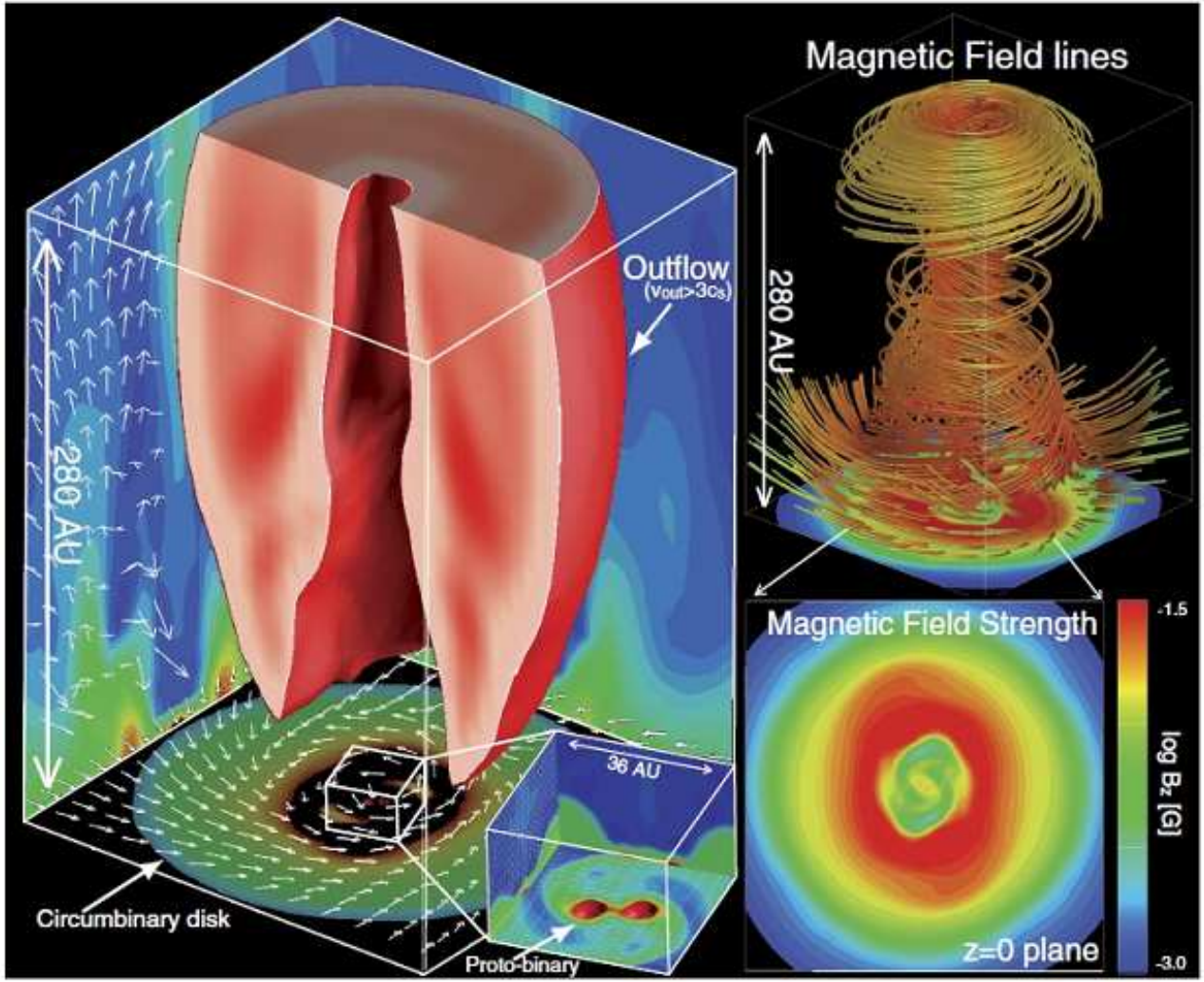}
\caption{
Large-scale structure of $l=10$ grid in three dimension.
{\it Left panel}: The configuration of circumbinary outflow is shown by red-volume, in which color means the outflow speed.
The density contours ({\it colors}) and velocity vectors ({\it arrows}) are projected in each wall surface.
The square at lower right corner shows the close-up view around the protobinary, in which color means the density distribution, and the high-density region (i.e., the protobinary) is represented by the red iso-density surface.
{\it Right upper panel}:
The magnetic field lines integrated from the circumbinary disk are plotted.
The bottom panel indicates the strength of $B_z$ by the color.
{\it Right lower panel}:
The strength of $B_z$ on $z=0$ plane is represented by the color.
}
\label{fig:3}
\end{figure}

Both in ideal and resistive models, the outflow appears $\sim100$\,yr after the first core formation.
This kind of outflow was already reported in many past studies \citep[e.g.,][]{tomisaka02,banerjee06,hennebelle08a}.
In Figures~\ref{fig:2}{\it k}-{\it o}, the outflowing region is represented by the white-dotted line inside which the gas is outflowing from the center of the cloud.
Outflow continues to driven near the first core in ideal model.
On the other hand, in resistive model, the driving point of the outflow moves outwardly with time as seen in Figures~\ref{fig:2}{\it k}-{\it o}, because the decoupled region of $Re<1$ expands outwardly.
In resistive model, the magnetic field is too weak to drive outflow in the region near the first core where the magnetic dissipation is effective.
Figure~\ref{fig:3} left panel shows the configuration of the outflow for resistive model, which indicates that the outflow driven by the circumbinary disk, not by the protostar (or proto-binary).
Figure~\ref{fig:3} upper right panel shows the configuration of magnetic field lines that are strongly twisted by the rotation of the circumbinary disk.
The circumbinary disk has a density of $n<10^{12}\cm$ and is well coupled with the magnetic field that can drive the outflow.
On the other hand,  the magnetic field is too weak to drive outflow in the region around protobinary.
Figure~\ref{fig:3} shows $B_z$ on the $z=0$ plane with the same scale of left panel, and indicates that the magnetic field is weak around protobinary but strong in the circumbinary disk.

\section{Discussion and Summary}
In this study, we show that the outflow can be driven by the circumbinary disk.
We calculated the evolution of outflow 1762\,yr after its emergence, in which outflow continues to be driven by the circumbinary disk and reach 427\,AU from the center of the cloud with a maximum speed of $8.4\km$.
At the end of the calculation, each fragment (i.e., each protobinary) has an only mass of $\sim0.03\msun$.
During calculation, the protobinary revolves about the center of the cloud (i.e., common center of gravity) $\sim20$ times, keeping a separation of about $\sim5-10$\,AU.
It is expected that the mass of the protobinary, and size and speed of outflow increase with time in the gas accretion phase.

Previous studies about the evolution of a rotating cloud showed that fragmentation frequently occurs only after the gas becomes optically thick (i.e., after the first core formation) and binary system is possible to form when the molecular cloud has a certain amount of the angular momentum  \citep{bodenheimer00,goodwin07}.
On the other hand, the study about a collapsing magnetized-rotating cloud showed that the magnetic field suppresses fragmentation and a single star tends to form in a strongly magnetized core, because the rotation (or angular momentum) to promote fragmentation is transferred by the magnetic effect such as the magnetic braking and outflow \citep{machida04,machida05b,price07,hennebelle08b}.
However, these studies adopted ideal MHD approximation.
In reality, the magnetic field dissipates in the high-density region in the collapsing cloud \citep{nakano02}.
Note that although the ambipolar diffusion and the Hall term effect can be also important for the magnetic dissipation, the Ohmic dissipation dominates in the high-density gas region \citep{machida07}.

We calculated the magnetized rotating cloud including the magnetic dissipation, and found the following: 
(1) fragmentation occurs to form the protobinary by the rapid rotation, since a weak magnetic field realized by the Ohmic dissipation makes the angular momentum transfer by the magnetic effect ineffective.
In addition, such weak field cannot drive outflow from the protobinary,
(2) strong outflow is driven by the circumbinary disk, because the magnetic field accumulates in the circumbinary disk where the removal and accretion of magnetic field are balanced, then the strong field is realized.
The binary system often shows a single outflow that may be driven by the circumbinary disk, not by one protostar in binary system.
For example, a single molecular outflow in L1551 IRS5, in which binary system is embedded,  may be driven by this mechanism.

\begin{figure}
\includegraphics[width=170mm]{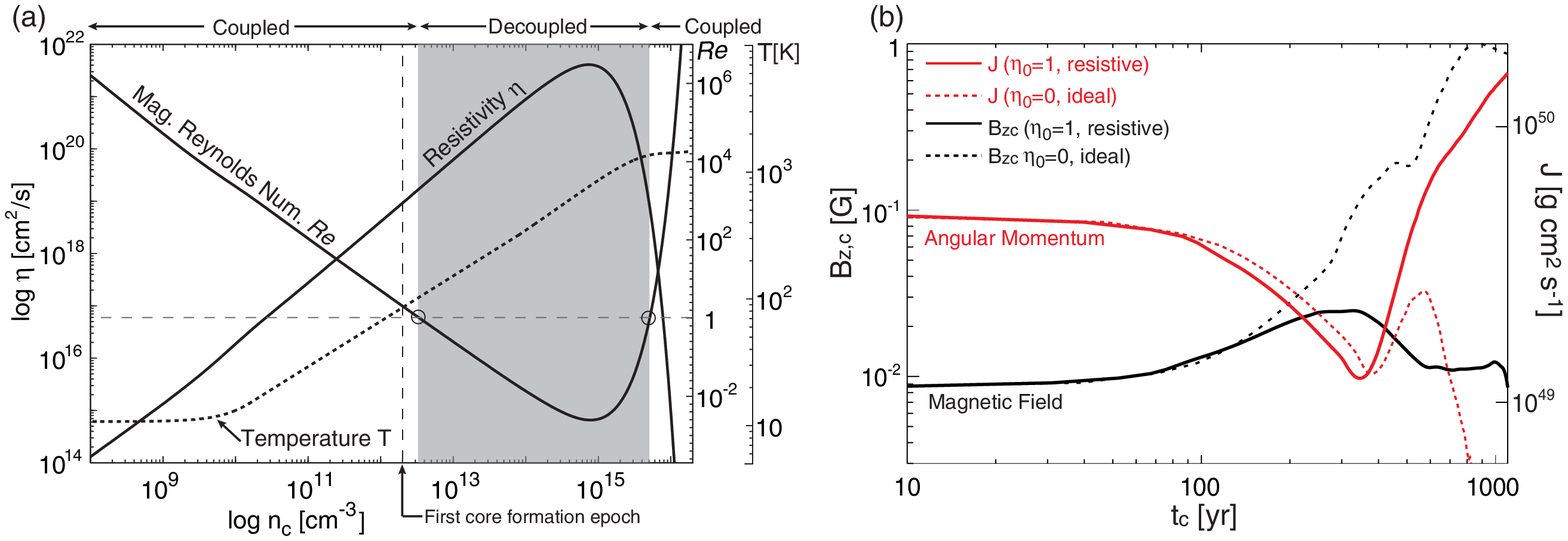}
\caption{
(a) Resistivity ($\eta$; left axis), magnetic Reynolds number ($Re$; right axis), and temperature ($T$; outer right axis) as a function of the number density.
The magnetic field well couple with the gas in ``coupled'' region, while magnetic field  decuples from the gas in ``decoupled'' region (grey zone).
The first core formation epoch is plotted by the arrow.
(b) The magnetic field strength $B_{\rm z,c}$ (left axis, black lines) at the center of the cloud, and angular momentum $J$ (right axis, red lines) in the region of $n > 0.1\,\nc$ against the elapsed time after the first core formation for ideal (dotted lines) and resistive (solid lines) models.
}
\label{fig:1}
\end{figure}

In the present study, we adopted the resistivity $\eta$ as most plausible value as shown in Figure~\ref{fig:1}{\it a}.
However, the resistivity depends on the size distribution and abundance of dusts.
The resistivity is closely related to the magnetic dissipation.
The magnetic field is related to most fundamental phenomena of the star formation: the angular momentum and magnetic flux problem, fragmentation and binary formation, and outflow driving. 
Thus, it is crucial to determine the resistivity for resolving major problems in star formation.
Since the strength and driving point of the outflow is directly related to the resistivity as shown in this paper, we can determine it with future higher-resolution observations such as ALMA.

\acknowledgments
Numerical computation in this work was carried out at the Yukawa Institute Computer Facility.
This work was supported by the Grant-in-Aid for the Global COE Program "The Next Generation of Physics, Spun from Universality and Emergence" from the Ministry of Education, Culture, Sports, Science and Technology (MEXT) of Japan, and partially supported by the Grants-in-Aid from MEXT (18740104, 20540238, 21740136).

\clearpage

\clearpage

\end{document}